\documentclass[aps,12pt,amsmath,amssymb,superscriptaddress,showpacs,showkeys]{revtex4}

\usepackage{dcolumn}
\usepackage{graphicx}
\usepackage{textcomp}
\usepackage{epsfig,bm}
\newcommand{\vi}[1]{\mbox{\boldmath $#1$}}
\newcommand{\vis}[1]{\mbox{\boldmath ${\scriptstyle #1}$}}

\begin{document}
\title{Significance and properties of internucleon correlation functions}
\author{Y. Suzuki}
\affiliation{Department of Physics, and Graduate 
School of Science and Technology, Niigata University, Niigata
950-2181, Japan}
\email{suzuki@nt.sc.niigata-u.ac.jp}
\author{W. Horiuchi}%
\affiliation{Graduate School of Science and Technology, 
Niigata University, Niigata 950-2181, Japan}
\email{horiuchi@nt.sc.niigata-u.ac.jp}
\pacs{21.60.Jz, 21.30.-x, 21.60.De}
\keywords{Correlation functions; Ground state energy; Nucleon-nucleon interaction}

\begin{abstract}
We show that a nuclear Hamiltonian and a set of internucleon correlation 
functions is in a one-to-one correspondence. The correlation functions 
for $s$-shell nuclei interacting via the two-nucleon interaction of AV8$^\prime$ 
type are calculated to exhibit the importance of tensor correlations as 
well as short-range central correlation. The asymptotic behavior of the 
correlation functions is also discussed. 
\end{abstract}
\maketitle

\section{Introduction}

According to the Hohenberg-Kohn theorem~\cite{hk}, the ground 
state of an 
interacting electron gas in an external potential is a unique 
functional 
of the density. Together with a practical method of prescribing the
density~\cite{ks}, a density functional theory (DFT) 
has played a vital role for 
calculating the ground state energy of 
electron systems.
Whether or not the DFT can be applied to a nucleus 
which is a self-bound system is an important question. Since the 
nuclear Hamiltonian includes no single-particle external 
potential, it is obvious that 
the proof of the Hohenberg-Kohn theorem 
does not apply for the 
nuclear ground state even though 
the application of the DFT is justifiable. 

There are several papers which appear to support the DFT for 
nuclei~\cite{engel,barnea,gjb}. The arguments made in these 
papers all assume some sort of intrinsic density for which the DFT is
discussed. For example, the intrinsic density is formed by 
putting the center of mass motion in some potential well or by assuming 
a symmetry violating intrinsic state. In the former case one has to 
separate the genuine internal motion from the center of mass motion, 
which is in general not trivial as in the case of large space 
shell model calculation encompassing major shell mixing.   
In the latter case a physical 
wave function is obtained by restoring the symmetry by a projection 
procedure. This approach has however only a limited validity, that is,
assuming an intrinsic state is already an approximation to a  
many-body theory. An intrinsic shape, if it is meaningful at all,
should appear 
automatically from a theory which has no recourse to the existence 
of such an intrinsic shape~\cite{forest,be8}. 
The two-$\alpha$ 
cluster structure for $^8$Be just comes out from a calculation 
which involves
no such assumption~\cite{be8}. Our recent four-nucleon 
calculation~\cite{he4} has succeeded, without assuming a cluster ansatz, 
to show that some of the excited states in $^4$He have 
$3N+N$ cluster configuration. 

A generalization of the DFT is discussed by introducing 
a pair density~\cite{ziesche,hetenyi} as a key quantity to 
characterize the system of interacting 
many particles.
The pair 
density or two-particle density gives a deeper insight into 
the internal structure of the system, especially into the correlated 
motion. 
The purpose of this paper is to examine internucleon correlation 
functions ($i$CF) since the energy of the nuclear ground state is 
manifestly a 
functional of these functions. Following the Hohenberg-Kohn theorem, 
we can unambiguously prove that the nuclear interaction can be 
uniquely determined by the $i$CF, that is, the nuclear Hamiltonian 
and the $i$CF has a one-to-one correspondence. 
Examples of $i$CF are given for $s$-shell nuclei. They are calculated 
using accurate wave functions obtained with realistic 
interactions. 

Since the nucleon-nucleon interaction depends on the spins and isospins 
of the nucleons, we have to consider 
the $i$CF in different spin-isospin channels. We discuss the relation 
between the various terms of 
the nucleon-nucleon interaction and the $i$CF as well as the 
asymptotic forms of the $i$CF. Information on the $i$CF 
is expected to be important for a class of variational calculations 
which use correlated trial wave functions 
including correlation factors such as 
variational Monte Carlo~\cite{vmc}, coupled-cluster theory~\cite{bishop},
Fermi hypernetted chain theory~\cite{chain}, and cluster expansion 
method~\cite{alvioli} and for a many-body theory using a unitary 
transformation of the nucleon-nucleon interaction~\cite{ucom}.

\section{Energy as a functional of internucleon correlation functions}

\subsection{Definition of internucleon correlation functions}
A Hamiltonian for a nucleus with $N$ nucleons is taken as 
\begin{align}
H=K+V=\sum_{i=1}^N \frac{1}{2m}{\bm{p}_i^2}-\frac{1}{2Nm}{\bm{P}}^2
+\sum_{i<j} v_{ij},
\label{hamiltonian}
\end{align}
where $m$ is the nucleon mass, $\bm{P}=\sum_i{\bm{p}}_i$ is the total 
momentum, and the center of mass kinetic energy is subtracted 
so as to calculate the internal energy of the nucleus.

A nucleon-nucleon interaction $v_{ij}$ may be expressed as follows
\begin{equation}
v_{ij}=\sum_p v^{(p)}(r_{ij}){\cal O}^{(p)}_{ij},
\label{nnforce}
\end{equation}
where $r_{ij}=|{\vi{r}}_{ij}|$ with ${\bm{r}}_{ij}={\bm{r}}_i-{\bm{r}}_j$ 
being
the relative distance of 
nucleons $i$ and $j$. Three-body forces are ignored for the sake of 
simplicity. The operators 
${\cal O}^{(p)}_{ij}$ denote various terms of the nucleon-nucleon 
potential. 
For the first eight terms, e.g., they are defined as
\begin{eqnarray}
& &{\cal O}^{(1)}_{ij}=1,\ \ \ {\cal O}^{(2)}_{ij}={\bm{\sigma}}_i\cdot{\bm{\sigma}}_j, 
\ \ \ {\cal O}^{(3)}_{ij}={\bm{\tau}}_i\cdot{\bm{\tau}}_j,\ \ \ 
{\cal O}^{(4)}_{ij}=({\bm{\sigma}}_i\cdot{\bm{\sigma}}_j)({\bm{\tau}}_i\cdot{\bm{\tau}}_j),
\nonumber \\
& &{\cal O}^{(5)}_{ij}=S_{ij},\ \ \ 
{\cal O}^{(6)}_{ij}=S_{ij}{\bm{\tau}}_i\cdot{\bm{\tau}}_j,\ \ \ 
{\cal O}^{(7)}_{ij}=({\bm{L}}\cdot{\bm{S}})_{ij},\ \ \ 
{\cal O}^{(8)}_{ij}=({\bm{L}}\cdot{\bm{S}})_{ij}{\bm{\tau}}_i\cdot{\bm{\tau}}_j,\label{av8}
\end{eqnarray}
where $S_{ij}=3(\hat{{\bm{r}}}_{ij}\cdot{\bm{\sigma}_i})
(\hat{{\bm{r}}}_{ij}\cdot{\bm{\sigma}_j})-{\bm{\sigma}_i}\cdot{\bm{\sigma}_j}$ is 
the tensor operator, and $({\bm{L}}\cdot{\bm{S}})_{ij}$ is the spin-orbit 
operator where ${\bm{L}}={\bm{r}_{ij}}\times {\bm{\pi}_{ij}}$ with ${\bm{\pi}_{ij}}
=-i({\partial}/{\partial {\bm{r}_{ij}}})$ and ${\bm{S}}=\frac{1}{2}({\bm{\sigma}_i}
+{\bm{\sigma}_j})$. The Coulomb potential is included in Eq.~(\ref{nnforce}) with 
$v^{(p)}(r_{ij})=e^2/r_{ij}$ and ${\cal O}^{(p)}_{ij}=P_{i\pi}P_{j\pi}$ where 
$P_{i\pi}$ is 1 for protons and 0 for neutrons. 
With the use of $\bm{p}_i-\bm{p}_j=2\hbar\bm{\pi}_{ij}$, 
the internal kinetic energy is rewritten as 
\begin{align}
K=\frac{1}{2Nm}\sum_{i<j}(\bm{p}_i-\bm{p}_j)^2=
\frac{2\hbar^2}{Nm}\sum_{i<j}\bm{\pi}_{ij}^2.
\end{align}

Let $\Psi$ and $E$ denote the wave function and energy of the 
ground state of the nucleus, that is, $H\Psi=E\Psi$.  The wave 
function  
$\Psi$ satisfies all of symmetry properties such as 
translation-invariance and rotation-invariance. Assuming 
that $\Psi$ is normalized, the energy is given as the expectation value 
of $H$, $E=\langle \Psi |H|\Psi \rangle$. Because $\Psi$ is 
antisymmetric with respect to an exchange of nucleons, $E$ is reduced to 
\begin{align}
E=\frac{N(N-1)}{2}\left[\frac{2\hbar^2}{Nm}
\langle \Psi|{\bm {\pi}}_{12}^2|\Psi\rangle 
+\langle \Psi|v_{12}|\Psi \rangle \right].
\end{align}
The expectation values of the kinetic and potential energy terms can be 
expressed in terms of the $i$CF. The kinetic energy term reads 
\begin{eqnarray}
\langle \Psi|{\bm {\pi}}_{12}^2|\Psi\rangle = \int d\bm{k}\, {\bm{k}^2} 
\langle \Psi|\delta({\bm {\pi}}_{12}-\bm{k})|\Psi\rangle 
=\int_0^{\infty} dk\, k^4 C(k),
\end{eqnarray}
where $\bm{k}=(k,\hat{\vi k})$ is just an integration variable, and 
$C(k)$ is the $i$CF in a momentum space, which is defined by
\begin{align}
C(k)=\langle \Psi|\frac{\delta({\pi}_{12}-k)}{k^2}|\Psi\rangle.
\end{align}
The calculation of  $C(k)$ is easily performed 
if the wave function $\Psi$ is given in the momentum space because 
$\bm{\pi}_{12}=(\pi_{12}, \hat{\vi \pi}_{12})$ is then just a multiplying operator. 

In exactly the same way, we can express the potential energy term as 
\begin{eqnarray}
\langle \Psi|v_{12}|\Psi \rangle&=&\sum_p \int  d\bm{r}\, v^{(p)}(r)
\langle \Psi|\delta({\bm {r}}_{12}-\bm{r}){\cal O}^{(p)}_{12}|\Psi\rangle
\nonumber \\
&=&\sum_p \int_0^{\infty}dr\, r^2 v^{(p)}(r) C^{(p)}(r),
\label{pot.contri}
\end{eqnarray}
where $C^{(p)}(r)$ is the $i$CF corresponding to the operator of type $p$ 
in the nucleon-nucleon potential
\begin{equation}
C^{(p)}(r)=\langle \Psi|\frac{\delta({r}_{12}-r)}{r^2}{\cal O}^{(p)}_{12}|\Psi\rangle.
\label{def.iCF}
\end{equation}
Here $\bm{r}=(r,\hat{\vi r})$ is not a dynamical coordinate but an integration variable. 
The $i$CF for the Coulomb potential is defined similarly
\begin{equation}
C_{\rm Coul}(r)=\langle \Psi|\frac{\delta({r}_{12}-r)}{r^2}P_{1\pi}P_{2\pi}|\Psi\rangle.
\end{equation}
The energy $E$ is thus manifestly a functional of several scalar 
$i$CF, $C(k)$ and $C^{(p)}(r)$: 
\begin{eqnarray}
E&=&(N-1)\frac{\hbar^2}{m}\int_0^{\infty} dk\,k^4 C(k) 
+\frac{N(N-1)}{2}\sum_p \int_0^{\infty}dr\, r^2 v^{(p)}(r) C^{(p)}(r).
\label{E_CF}
\end{eqnarray}

As seen above, 
the energy can be expressed in terms of $i$CF.
They are different from a two-nucleon density, which is defined as
\begin{eqnarray}
\rho(\bm{r},\bm{R})
&=&\langle \Psi
 |\delta(\bm{r}_1-\bm{x}_N-\bm{R}-\textstyle{\frac{1}{2}}\bm{r})
\delta(\bm{r}_2-\bm{x}_N-\bm{R}+\textstyle{\frac{1}{2}}\bm{r})
|\Psi \rangle
\nonumber \\
&=&\langle \Psi |\delta(\bm{r}_1-\bm{r}_2-\bm{r})
\delta(\textstyle{\frac{1}{2}}(\bm{r}_1+\bm{r}_2)-\bm{x}_N-\bm{R})
|\Psi \rangle,
\end{eqnarray}
where $\bm{x}_N$ is the center of mass coordinate of the nucleus. 
Because $v_{12}$ is independent of where the center of mass of the 
two nucleons relative to the total center of mass is located, 
i.e., of the coordinate $\frac{1}{2}(\bm{r}_1+\bm{r}_2)-\bm{x}_N$, 
in calculating the expectation value of the potential energy 
we can integrate over $\bm{R}$, that is, we only need 
\begin{align}
\int d{\bm{R}}\, \rho(\bm{r},\bm{R}),
\end{align}
which is nothing but the $i$CF, $C^{(1)}(\bm{r})$.  Note that 
$C^{(1)}(\bm{r})$ is different from an 
intrinsic one-body density, which is defined as  
\begin{align}
\rho^{(1)}(\bm r)=N\langle \Psi |\delta({\bm r}_1-{\bm x}_N-{\bm r})|\Psi \rangle.
\end{align}

\subsection{Extension of Hohenberg-Kohn theorem}

We think that 
no consensus has yet been 
reached on 
the existence of DFT 
for the nuclear Hamiltonian (\ref{hamiltonian}). 
We can prove, however, that the $i$CF  
can constitute a set of basic variables for the nuclear system. 
Obviously $C^{(p)}(r)$ 
for the ground state are uniquely determined by $v_{ij}$, and hence they 
are functionals of $v_{ij}$. 
Following the proof used in Ref.~\cite{hk}, we can prove that, conversely, 
$v_{ij}$ is a unique functional 
of $C^{(p)}(r)$. 
For this purpose we only need to show that 
$v_{ij}$ is uniquely determined by $C^{(p)}(r)$. 
Let us assume that the ground state of the 
Hamiltonian (\ref{hamiltonian}) is non-degenerate. 
Assume 
that, contrary to the statement to be proved, there is
another potential $v_{ij}'$, which gives rise to a ground 
state wave function $\Psi'$ and an energy $E'$, 
resulting from
the same $i$CF $C^{(p)}(r)$.  Clearly $\Psi'$ cannot 
be equal to $\Psi$, because they satisfy different Schr\"odinger 
equations. Let $H'$ denote 
the Hamiltonian obtained by replacing $v_{ij}$ with $v_{ij}'$. 
Then, from the Ritz theorem, we have that
\begin{align}
E'=\langle \Psi'|H'|\Psi'\rangle < \langle \Psi|H'|\Psi \rangle
=\langle \Psi|H+V'-V|\Psi \rangle.
\label{ritz.th}
\end{align}
Here the inequality 
$\langle \Psi'|H'|\Psi'\rangle < \langle \Psi|H'|\Psi \rangle$ holds 
because $\Psi$ is different from $\Psi'$. Using Eq.~(\ref{E_CF}) leads to 
\begin{eqnarray}
E'&<& E + \frac{N(N-1)}{2}\sum_{p} \int_0^{\infty} dr\,  r^2
[v'^{(p)}(r)-v^{(p)}(r)]C^{(p)}(r).
\label{hk1}
\end{eqnarray}
Interchanging primed and unprimed quantities yields the result 
\begin{eqnarray}
E&<& E' + \frac{N(N-1)}{2}\sum_{p}\int_0^{\infty} dr\, r^2 
[v^{(p)}(r)-v'^{(p)}(r)]C^{(p)}(r).
\label{hk2}
\end{eqnarray}
Adding up 
Eqs.~(\ref{hk1}) and (\ref{hk2}) leads to the well-known 
inconsistency 
\begin{align}
E+E' < E+E'.
\end{align}
Thus we can conclude that $v_{ij}$ is a unique functional of 
$C^{(p)}(r)$. Since $v_{ij}$ 
specifies $H$ unambiguously, 
it is concluded that 
the nuclear ground state is a unique functional of $C^{(p)}(r)$. 
The ground state energy $E$ takes a minimum for the 
exact $i$CF. 

The wave function $\Psi$ depends on $3N-3$ variables as well as 
the spin and isospin coordinates. It is therefore hopeless to 
obtain an accurate wave function for 
$N\gtrsim 10$ 
using a basis expansion 
method. Contrary to this  
approach, the above consideration tells us that to calculate 
the ground state energy accurately 
we need to know about 10-20 $i$CF  
which are all single-variable scalar functions. 
It is interesting to know   
the characteristic behaviors, e.g., the shapes and magnitudes of these CF.  

The $i$CF satisfy the following equations
\begin{eqnarray}
& &\int_0^{\infty}dk\, k^2 C(k)=1,
\nonumber \\
& &\int_0^{\infty}dr\, r^2 C^{(p)}(r) =
\frac{2}{N(N-1)}
\langle \Psi|\sum_{i<j}{\cal O}^{(p)}_{ij}|\Psi \rangle.
\label{norm.C}
\end{eqnarray}
Using the identity
\begin{equation}
\sum_{i<j}({\vi r}_i-{\vi r}_j)^2=N\sum_{i=1}^N({\vi r}_i-{\vi x}_A)^2,
\end{equation}
the root mean square matter radius of the nucleus can be calculated 
from a moment of $C^{(1)}(r)$ as 
\begin{equation}
\Big\langle \frac{1}{N}\sum_{i=1}^N({\vi r}_i-{\vi x}_N)^2\Big\rangle
=\frac{N-1}{2N}\int_0^{\infty}dr \,r^4C^{(1)}(r). 
\label{rms}
\end{equation}

An interesting relation is obtained by expressing the left 
side of Eq.~(\ref{rms}) with the use of 
the  one-particle density 
\begin{equation}
\int d{\vi r}\, r^2 \rho^{(1)}({\vi r})=\frac{N-1}{2}\int_0^{\infty}
 dr\, r^4C^{(1)}(r).
\end{equation}
For a spherical density we have the following relation
\begin{equation}
\int_0^{\infty} dr\, r^4 \rho^{(1)}({r})=
\frac{N-1}{2}\int_0^{\infty}dr \,r^4C^{(1)}(r)
\label{C.rho.relation}
\end{equation}
with
\begin{equation}
\rho^{(1)}(r)=\int d\hat{\vi r} \, \rho^{(1)}({\vi r})
=N\langle \Psi|\frac{\delta(|{\vi r_1}-{\vi x}_N|-r)}{r^2}|\Psi \rangle.
\end{equation}

\subsection{Internucleon correlation functions in spin and isospin channels}
The characteristics of nucleon-nucleon 
potentials may be more transparent if we 
decompose them into four spin and isospin channels 
of two nucleons, $(ST)=(10),\, (01),\, (11),\, (00)$, instead of 
using the operator representation of Eqs.~(\ref{nnforce}) and 
(\ref{av8}). To do this we
use the following identities 
\begin{eqnarray}
& &1=P^{(10)}_{ij}+P^{(01)}_{ij}+P^{(11)}_{ij}+P^{(00)}_{ij},\ \ \ 
{\vi \sigma}_1\cdot{\vi \sigma_2}=P^{(10)}_{ij}-3P^{(01)}_{ij}+P^{(11)}_{ij}-3P^{(00)}_{ij},
\nonumber \\
& &{\vi \tau}_1\cdot{\vi \tau_2}=-3P^{(10)}_{ij}+P^{(01)}_{ij}+P^{(11)}_{ij}-3P^{(00)}_{ij},\nonumber \\
& &{\vi \sigma}_1\cdot{\vi \sigma_2}{\vi \tau}_1\cdot{\vi
 \tau_2}=-3P^{(10)}_{ij}-3P^{(01)}_{ij}+P^{(11)}_{ij}+9P^{(00)}_{ij},
\end{eqnarray}
where  
$P^{(ST)}_{ij}$ is the projection operator which 
projects onto the state with $(ST)$ of the two nucleons $i$ and $j$.
The nucleon-nucleon potential (\ref{nnforce}) is expressed as 
\begin{equation}
v_{ij}=\sum_{(ST)\cal O}v^{(ST)}_{\cal O}(r_{ij}){\cal O}_{ij} 
P^{(ST)}_{ij},
\label{pot_ST}
\end{equation}
where the summation label $\cal O$ indicates summing over 
various components of the nucleon-nucleon interaction. 
For example, they stand for 
central (1), tensor (T), and spin-orbit (LS), and the corresponding 
operators ${\cal O}_{ij}$ denote 1, $S_{ij}$, and 
$({\vi L}\cdot{\vi S})_{ij}$, respectively. 
For the potential (\ref{nnforce}), the 
form factor $v^{(ST)}_{\cal O}$ is related to those of $v^{(p)}$ as 
follows:
\begin{eqnarray}
& &v_1^{(10)}=v^{(1)}+v^{(2)}-3v^{(3)}-3v^{(4)},
\ \ \ v_{\rm T}^{(10)}=v^{(5)}-3v^{(6)},\ \ \ 
v_{\rm LS}^{(10)}=v^{(7)}-3v^{(8)},\nonumber \\
& &v_1^{(01)}=v^{(1)}-3v^{(2)}+v^{(3)}-3v^{(4)},\ \ \ 
v_{\rm T}^{(01)}=v^{(5)}+v^{(6)},\ \ \ 
v_{\rm LS}^{(01)}=v^{(7)}+v^{(8)},\nonumber \\
& &v_1^{(11)}=v^{(1)}+v^{(2)}+v^{(3)}+v^{(4)},\ \ \ 
v_{\rm T}^{(11)}=v^{(5)}+v^{(6)},\ \ \ 
v_{\rm LS}^{(11)}=v^{(7)}+v^{(8)},\nonumber \\
& &v_1^{(00)}=v^{(1)}-3v^{(2)}-3v^{(3)}+9v^{(4)},\ \ \ 
v_{\rm T}^{(00)}=v^{(5)}-3v^{(6)},\ \ \ 
v_{\rm LS}^{(00)}=v^{(7)}-3v^{(8)}.
\end{eqnarray}

The $i$CF terms corresponding to the potential form of
Eq.~(\ref{pot_ST}) are
\begin{align}
C_{\cal O}^{(ST)}(r)=
\langle\Psi |\frac{\delta(r_{12}-r)}{r^2}{\cal O}_{12}P^{(ST)}_{12}|\Psi \rangle. 
\label{dens.2}
\end{align}
The relationship between the two $i$CF, 
Eqs.~(\ref{def.iCF}) and (\ref{dens.2}), reads 
\begin{eqnarray}
& &C_{1}^{(10)}=\frac{1}{16}(3C^{(1)}+C^{(2)}-3C^{(3)}-C^{(4)}), \ \ \ 
C_{\rm T}^{(10)}=\frac{1}{4}(C^{(5)}-C^{(6)}), \nonumber \\
& &C_{\rm LS}^{(10)}=\frac{1}{4}(C^{(7)}-C^{(8)}),\nonumber \\
& &C_{1}^{(01)}=\frac{1}{16}(3C^{(1)}-3C^{(2)}+C^{(3)}-C^{(4)}),\ \ \ 
C_{\rm T}^{(01)}=0, \ \ \ C_{\rm LS}^{(01)}=0,\nonumber \\
& &C_{1}^{(11)}=\frac{1}{16}(9C^{(1)}+3C^{(2)}+3C^{(3)}+C^{(4)}), \ \ \ 
C_{\rm T}^{(11)}=\frac{1}{4}(3C^{(5)}+C^{(6)}), \nonumber \\
& &C_{\rm LS}^{(11)}=\frac{1}{4}(3C^{(7)}+C^{(8)}),\nonumber \\
& &C_{1}^{(00)}=\frac{1}{16}(C^{(1)}-C^{(2)}-C^{(3)}+C^{(4)}),\ \ \ 
C_{\rm T}^{(00)}=0, \ \ \ C_{\rm LS}^{(00)}=0.
\end{eqnarray}
Equation~(\ref{E_CF}) is rewritten using these $i$CF as follows 
\begin{eqnarray}
E&=&(N-1)\frac{\hbar^2}{m}\int_0^{\infty} dk\,k^4 C(k) 
+\frac{N(N-1)}{2}\sum_{(ST)\cal O} \int_0^{\infty}dr\, r^2 
v^{(ST)}_{\cal O}(r) C^{(ST)}_{\cal O}(r).
\end{eqnarray}

\section{Specific examples}

Recently we have developed a method of calculating matrix elements 
for the interaction of Eq.~(\ref{av8}) as well as various types of 
$i$CF using correlated Gaussian functions 
with the orbital motion being described in two global vectors~\cite{fbs}.  
The accuracy of the formulation has been tested by comparing to 
other calculations for $N=3-4$ nuclei. An application of the 
method to 
studying excited states of $^4$He has met a fair success, revealing 
an inversion doublet picture arising from  
$^3$H($t$)+$p$ and $^3$He($h$)+$n$ cluster structure~\cite{he4}. 

The wave functions are expressed as a combination of 
many basis states, each of which  
has the following $LS$ coupling form    
\begin{equation}
\Psi_{(LS)JM_JTM_T}={\cal A}[\psi_L^{(\rm space)}\psi_S^{(\rm spin)}]_{JM_J}
\psi_{TM_T}^{(\rm isospin)},
\label{basis.LS}
\end{equation}
where the square bracket $[\ldots]$ stands for the angular momentum
coupling, and the antisymmetry 
of nucleons is met by the antisymmetrizer ${\cal A}$. 
The spin and isospin parts are expanded using the basis 
of successive coupling, e.g.,  
\begin{equation}
\psi_{SM_S}^{(\rm spin)}=\big|[\cdots [[[\textstyle{\frac{1}{2}}
\textstyle{\frac{1}{2}}]_{S_{12}}
\textstyle{\frac{1}{2}}]_{S_{123}}]\cdots]_{SM_S}\rangle,
\end{equation}
where 
the set of intermediate spins $(S_{12},S_{123},\ldots)$ is allowed 
to take all possible values for a given $S$.

\begin{figure}[b]
\epsfig{file=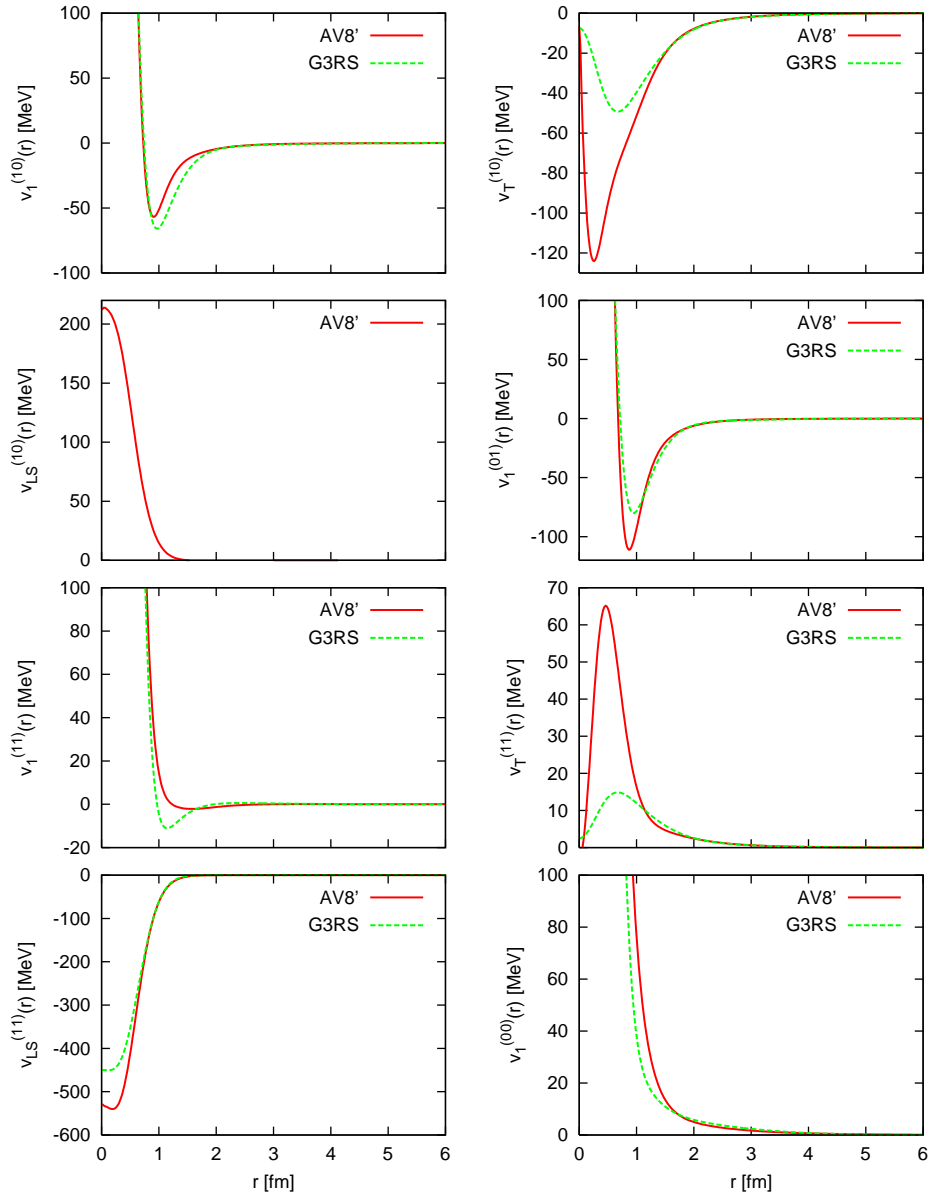,scale=1.0}
\caption{The nucleon-nucleon potentials of AV8$^\prime$~\cite{av8} and 
G3RS~\cite{tamagaki}.}
\label{av8-g3rs}
\end{figure}

The orbital part $\psi_{LM}^{(\rm space)}$ is given as follows 
\begin{equation}
F_{(L_1L_2)LM}(u_1,u_2,A,{\vi x})
= {\rm exp}\left(-{\frac{1}{2}}{\widetilde{{\vi x}}} A {\vi x}\right)
[{\cal Y}_{L_1}({\widetilde{u_1}}{\vi x}) 
{\cal Y}_{L_2}({\widetilde{u_2}}{\vi x})]_{LM},
\label{cg}
\end{equation}
with ${\cal Y}_{LM}({\vi r})=r^LY_{LM}(\hat{\vi r})$, 
where ${\vi x}=({\vi x}_1,{\vi x}_2,\ldots,{\vi x}_{N-1})$ 
is a set of relative coordinates, say the Jacobi 
coordinate set, and $u_1$ and $u_2$ are $(N-1)$-dimensional 
column vectors which define the global vectors. Here 
${\widetilde{{\vi x}}} A {\vi x}=\sum_{i,j=1}^{N-1}A_{ij}{\vi
x}_i\cdot{\vi x}_j$ with $A_{ij}=A_{ji}$ and ${\widetilde{u_1}}{\vi
x}=\sum_{i=1}^{N-1}{u_1}_i{\vi x}_i$.
As we see, 
each basis function is characterized by a set of parameters, 
$A=(A_{ij})$, $u_1$, $u_2$, $L_1, L_2, L, S_{12},
S_{123},\ldots, S, T_{12}, T_{123},\ldots$.

The calculation of Hamiltonian matrix elements is made possible  
with the aid of the generating function $g$ 
\begin{equation}
g({\vi s}; A, {\vi x})=\exp\Big(-{\frac{1}{2}}{\widetilde{\vi x}}A{\vi x}+
\widetilde{\vi s}{\vi x}\Big),
\label{DEFGFN}
\end{equation}
where ${\vi s}$ is an $(N-1)$-dimensional column 
vector whose $i$th element is a 3-dimensional vector ${\vi s}_i$. 
By expressing ${\vi s}_i$ 
with 3-dimensional unit vectors ${\vi e}_1$ and ${\vi e}_2$ as 
${\vi s}_i={\lambda}_1{\vi e}_1 u_{1_i}+{\lambda}_2{\vi e}_2 u_{2_i}$, 
the basis function (\ref{cg}) is generated as follows:
\begin{eqnarray}
&& F_{(L_1L_2)LM}(u_1,u_2,A,{\vi x}) 
= {\frac{B_{L_1}B_{L_2}}{ L_1!L_2!}}\int\!\int 
d{\vi e}_1\, d{\vi e}_2 \, [Y_{L_1}({\vi e}_1) 
Y_{L_2}({\vi e}_2)]_{LM} \nonumber \\
& &\qquad \qquad \qquad \qquad \times\,  {\frac{\partial^{L_1+L_2}}
 {\partial{\lambda}_1^{L_1}\partial{\lambda}_2^{L_2}}}\, 
g({\lambda}_1{\vi e}_1 u_1\!+\!{\lambda}_2{\vi e}_2 u_2; A,{\vi x})
\Big\vert_{\lambda_1=\lambda_2=0}, 
\label{gfn}
\end{eqnarray}
where  
\begin{equation}
B_L={\frac{(2L+1)!!}{ 4 \pi}}.
\end{equation} 
Formulas for the matrix elements are given in Ref.~\cite{fbs}.  
In Appendix~\ref{appendix.a} we give 
a formula to calculate the $i$CF 
for the spin-orbit force, 
which was not included
in Ref.~\cite{fbs}.

We study the $i$CF of $s$-shell nuclei, $d, \,t, \,h$, and  
$^4$He ($\alpha$). We also show the $i$CF calculated for 
the first excited $0^+$ state of $^4$He, which is called 
$\alpha^*$ in this paper. Although this state 
decays into the $t+p$ channel with a width of 0.50\,MeV, 
approximating it  
as a bound state is fairly good, and the $i$CF 
of $\alpha^*$ are calculated using the wave function obtained 
in that approximation~\cite{he4}. Because $\alpha^*$ is a spatially 
extended state with a cluster 
structure of $t+p$ and $h+n$, comparing the $i$CF between 
$\alpha^*$ and other cases reveals how much 
the shapes and magnitudes of the $i$CF are modified by 
the structures
of the underlying nuclear states.

We use the AV8$^\prime$ potential~\cite{av8} and 
the G3RS potential~\cite{tamagaki} as the two-nucleon interaction. 
Both of them contain central, tensor and spin-orbit terms. 
The ${\bm{L}}^2$ and $({\bm{L}\cdot{\bm S}})^2$
terms of the G3RS potential are ignored. 
The radial form factors $v^{(ST)}_{\cal O}$ of the two potentials 
are displayed in Fig.~\ref{av8-g3rs}. As is well-known, the
longest-range attraction of the two-nucleon 
interaction is that belonging to 
the $v_{\rm T}^{(10)}$ term. In the 
intermediate region $(1<r<1.6)$\,fm, the singlet even central potential 
$v^{(01)}_1$ is most attractive, and then the $v_{\rm T}^{(10)}$ and 
$v_1^{(01)}$ terms follow. The central potentials all have strong 
short-range repulsion. Generally speaking, the G3RS potential is 
softer than the AV8$^\prime$ potential. The tensor force of the latter potential 
is much stronger at $r<1\,$fm than that of the G3RS potential. The G3RS 
potential has no $v_{\rm LS}^{(10)}$ component. 
In spite of these differences, the two
potentials give rather similar binding energies for the $s$-shell 
nuclei~\cite{fbs}. 

\begin{figure}[b]
\epsfig{file=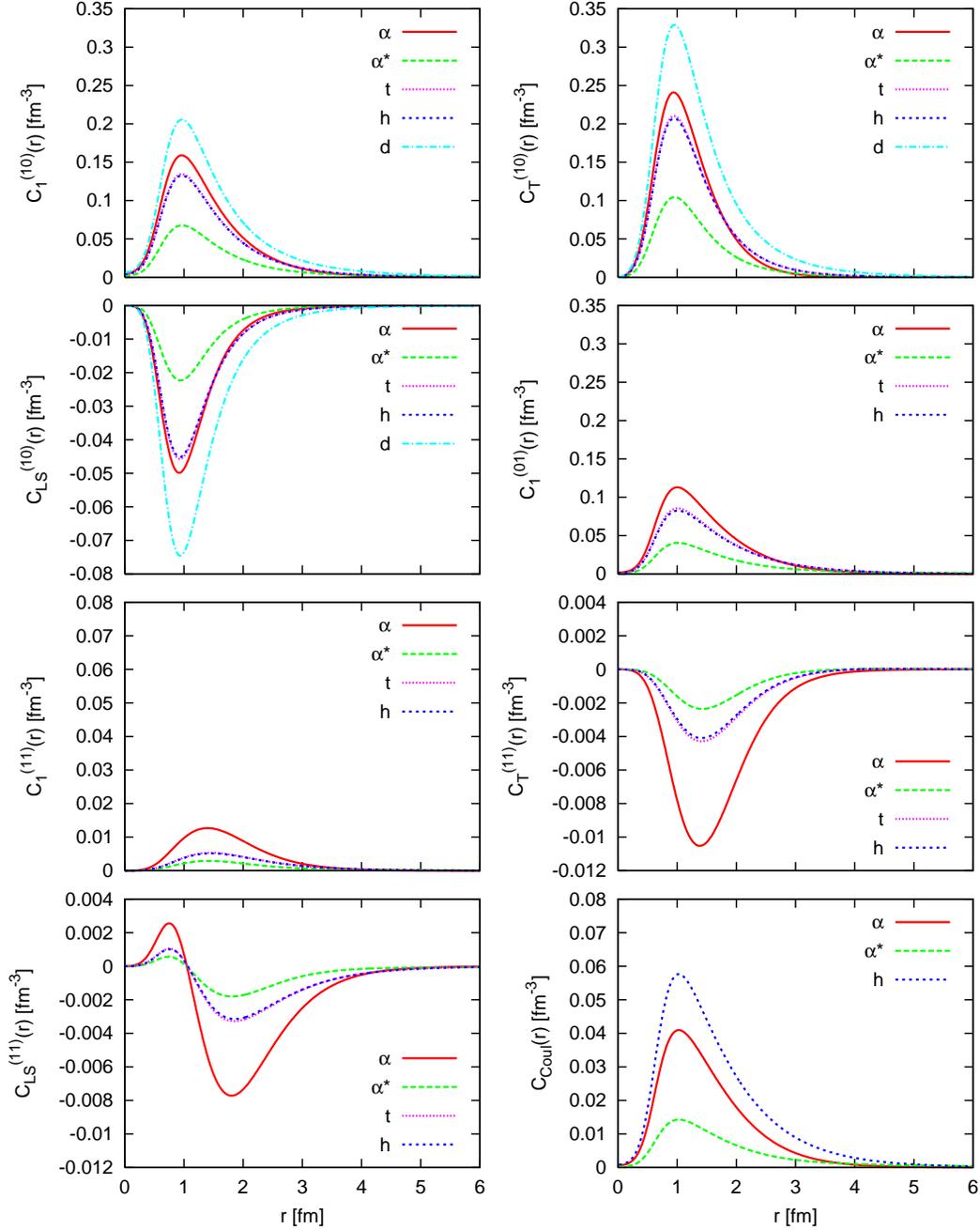,scale=1.1}
\caption{The $i$CF for $\alpha,\,
 \alpha^*$ (the first excited $0^+$ state of $^4$He), $t,\, h,\, d$ calculated 
using the AV8$^\prime$ potential. Note that different scales are used 
for the vertical axes.}
\label{av8.cf}
\end{figure}

\begin{figure}[b]
\epsfig{file=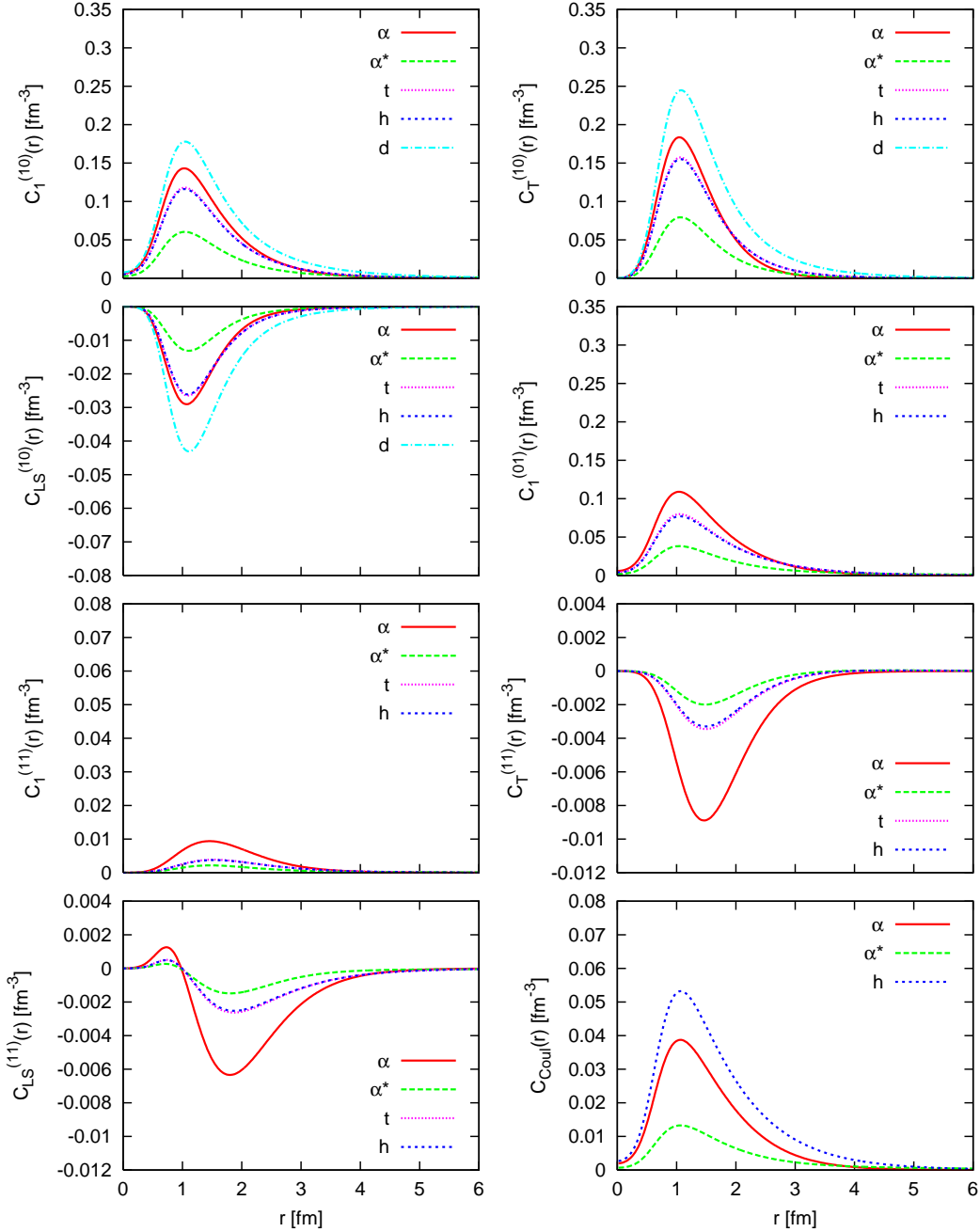,scale=1.1}
\caption{The same as Fig.~\ref{av8.cf} but calculated 
using the G3RS potential.} 
\label{g3rs.cf}
\end{figure}

Figures~\ref{av8.cf} and \ref{g3rs.cf} compare the $i$CF, 
$C^{(ST)}_{\cal O}(r)$, 
calculated using the wave functions obtained with 
the AV8$^\prime$ Hamiltonian with those obtained with the G3RS Hamiltonian.  
The amplitude of $C_1^{(00)}$ is by far smaller than the
others because of the purely repulsive nature of the 
corresponding potential $v_1^{(00)}$, and thus it is not shown in the 
figures. 
Common to the two cases is that the three $i$CF, 
$C_{\rm T}^{(10)},\,C_1^{(10)}$ and $C_1^{(01)}$, among others have much
larger amplitudes than the others, all having a peak at 
around 1\,fm. This is understandable from the characteristics 
of the $v^{(ST)}_{\cal O}$ curves shown 
in Fig.~\ref{av8-g3rs}. The more 
attractive the potential component, the larger the amplitude of the 
corresponding $i$CF. The central potential for even partial waves has a 
minimum around 1\,fm. Since the central and tensor forces in the triplet 
even channels couple, the $C_{\rm T}^{(10)}$ curve also has 
a peak at almost
the same position as that of $C_1^{(10)}$,  
though the $v_{\rm T}^{(10)}$ potential does not have a minimum around 1\,fm. 
Both of $C_1^{(11)}$ and $C_{\rm T}^{(11)}$ curves have a peak around
1.4-1.5\,fm because the central and tensor forces 
in the $(ST)=(11)$ channel 
couple and the central force has a shallow attraction beyond 1\,fm.
The peak position of the $i$CF for the Coulomb potential 
is the same as that of the main terms of the triplet even channel.

Because of 
the strong short-range repulsion of the central potentials $v_{1}^{(ST)}$, 
all the $C^{(ST)}_{\cal O}$ vanish near the origin. 
The spin-orbit force ($v^{(11)}_{\rm LS}$) has a very strong attraction 
around 1\,fm but it is confined to the short distance 
region. Comparing Figs.~\ref{av8.cf} and \ref{g3rs.cf}, the difference 
of the $i$CF produced by the potential models is mild in spite of 
apparently different behaviors of the potential form factors, for example 
of the tensor terms. 

The $s$-shell nuclei all produce 
similar shapes for each of the $i$CF. 
It is interesting to note the $i$CF of $\alpha^*$ exhibit the patterns
similar to the other cases, despite the fact that its structure is quite 
different from that of $\alpha$. Of course the amplitudes at larger values of
$r$ are much larger for $\alpha^*$ than for the other cases.

A remarkable characteristics of the $i$CF is that their asymptotics  
for a given nucleus are the same for all the $C^{(ST)}_{\cal O}$.   
For example, the eight 
$i$CF calculated for $\alpha$ using the AV8$^\prime$ potential follow 
$\sim \exp(-2\kappa r)/r^{2.09}$ with $\kappa=0.75$\,fm$^{-1}$ 
for $r$ larger than 6\,fm.

The case of the deuteron is easily understood. 
The deuteron wave function consists of the $S$- and $D$-wave components 
\begin{equation}
\Psi=\phi_0(r)Y_{00}(\hat{\vi r})\chi_{1M}\eta_{00}+\phi_2(r)[Y_2(\hat{\vi r})\chi_1]_{1M}\eta_{00},
\end{equation}
where $\chi_1$ and $\eta_0$ are the spin and isospin functions of the 
deuteron. 
For large values of $r$ for which the nuclear potential 
between the two nucleons in the deuteron are negligible, 
the Hamiltonian for the deuteron reduces to 
the kinetic energy alone, and thus the radial function $\phi_{\ell}(r)$ 
should be given by a solution of the 
free-particle Schr\"odinger equation with 
the negative energy ($-{\hbar^2\kappa^2}/m$) of the deuteron, that is the 
spherical Hankel function 
of the first kind $h_{\ell}^{(1)}(i\kappa r)$.  
The asymptotic form of the deuteron wave function is therefore given by 
\begin{equation}
\Psi = \sum_{\ell=0,2}K_{\ell}h_{\ell}^{(1)}(i\kappa r)
[Y_{\ell}(\hat{\vi r})\chi_1]_{1M}\eta_{00}
\end{equation}
with suitable coefficients $K_{\ell}$. The $i$CF of the 
deuteron for large $r$ reduces to 
\begin{equation}
C^{(ST)}_{\cal O}(r) \sim \sum_{\ell, \ell'=0,2} K_{\ell}K^*_{\ell'}\langle [Y_{\ell'}\chi_1]_{1M}\eta_{00}
|{\cal O}P^{(ST)}|[Y_{\ell}\chi_1]_{1M}\eta_{00} \rangle 
h_{\ell}^{(1)}(i\kappa r){h_{\ell'}^{(1)}}^*(i\kappa r).
\end{equation}
All of the $i$Cf for the deuteron displayed in Figs.~\ref{av8.cf} and 
\ref{g3rs.cf} satisfy the above behavior for $r \ge 5\,$fm.

The asymptotic behavior for other cases is discussed in 
Appendix~\ref{appendix.b} by taking into account the Coulomb force.
The case of $\alpha$ is understood by taking the nucleus 
R (in the notation of Appendix~\ref{appendix.b}) as the $pn$ system, 
which gives $Z=2$. See Eq.~(\ref{def.charge}). As discussed in 
Appendix~\ref{appendix.b}, the asymptotic behavior 
is given by $r^{-2-2\eta}\exp(-2\kappa r)$, where $\eta$ defined in  
Eq.~(\ref{eta.kappa}) becomes 0.046 for $\kappa=0.75$\,fm$^{-1}$. Thus 
we can understand the asymptotic behavior of the $i$CF noted above.

\begin{figure}[t]
\epsfig{file=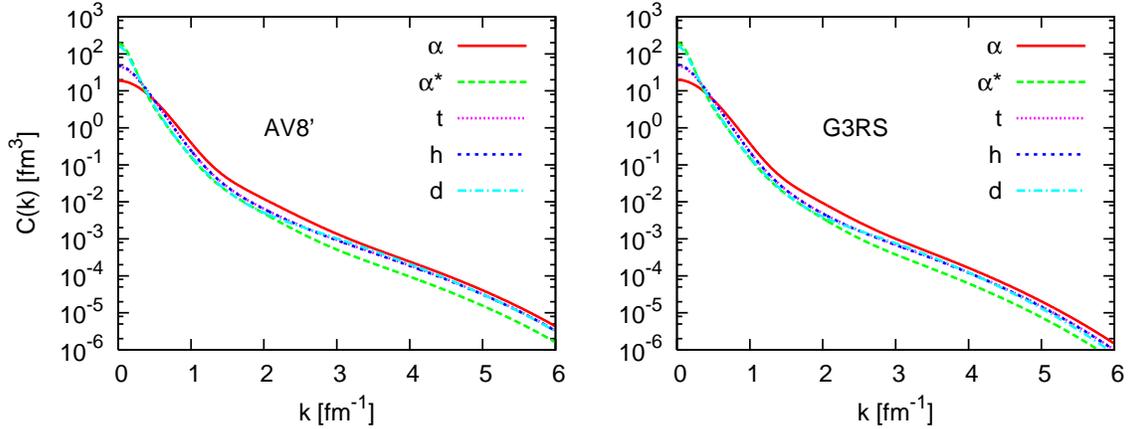,scale=1.3}
\caption{The $i$CF for the kinetic energy operator.} 
\label{kine.cf}
\end{figure}

The $i$CF $C(k)$ for the kinetic energy operator is shown in 
Fig.~\ref{kine.cf}. 
The AV8$^\prime$ and G3RS potentials give
qualitatively very similar results.  
The behavior of $C(k)$ for small values of $k$ is given analytically as 
explained in Appendix~\ref{appendix.b}. The numerical results 
confirm that 
the asymptotic form, Eq.~(\ref{asym.c(k)}), agrees with $C(k)$ 
for small values of $k$. 
The behavior of $C(k)$ for large values of $k$ 
primarily reflects the short-range central correlation and the tensor
correlation involved in the wave function. The enhancement of the 
curve around $k\sim 1.3$\,fm$^{-1}$ is due to the tensor force, as shown 
in Ref.~\cite{he6}.

\begin{figure}[b]
\epsfig{file=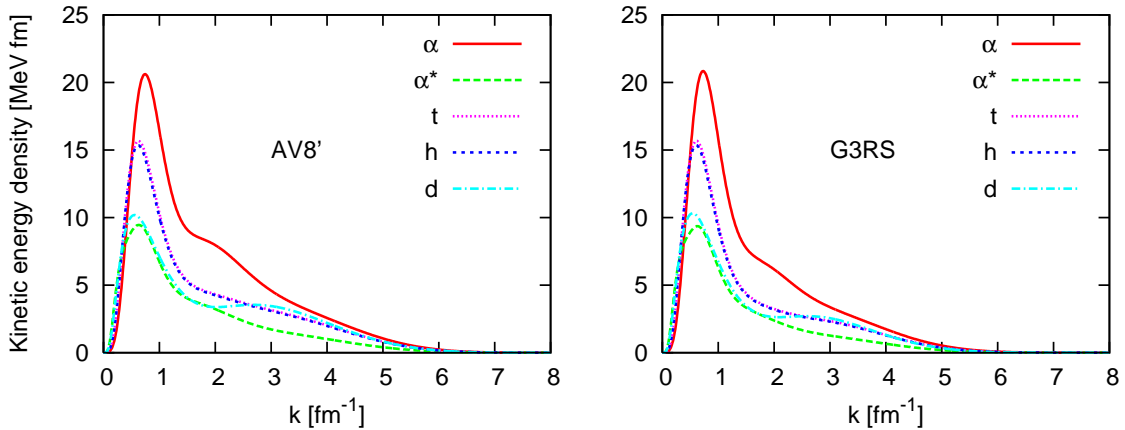,scale=1.3}
\caption{Two-nucleon relative kinetic energy density per unit wave number.}
\label{kine.density}
\end{figure}

Figure~\ref{kine.density} displays the kinetic energy density of the 
two-nucleon relative motion, $(\hbar^2/m)k^4C(k)$, as a function of $k$.
We clearly see that the tensor correlation increases the kinetic energy 
density beyond $k=1$\,fm$^{-1}$. The height of the density around
$k=1.5$\,fm$^{-1}$ is related to the components of higher partial waves 
induced by the tensor force. The $\alpha$ particle contains 
the largest bump among 
the $s$-shell nuclei, but the component contained in the 
first excited $0^+$ state ($\alpha^*$) is much 
smaller as understood from 
the $3N+N$ cluster structure. The kinetic energy density  
extends beyond 
$k=6$\,fm$^{-1}$ for all the nuclei, which is of course due to the 
short-range repulsion of the central interaction.  

\section{Summary}

We have shown that the ground state energy of a nucleus as a self-bound
system is a functional of a set of internucleon correlation
functions including the kinetic energy term. Conversely the set of 
the internucleon correlation
functions uniquely determines the nuclear Hamiltonian. 
Namely, there is a one-to-one correspondence between the
nuclear Hamiltonian and the internucleon correlation functions. 
The ground state energy becomes a minimum for a set of the exact
internucleon correlation functions.  

Using the accurate wave functions for $s$-shell nuclei, we have
calculated the internucleon correlation functions for $d,\, ^3$H, $^3$He
and the ground state of $^4$He. To see the dependence of the correlation 
functions on nuclear structure, we have also included the first 
excited $0^+$ state of $^4$He. We used two different potentials, AV8$^\prime$
and G3RS, as the two-nucleon interaction. Both of them contain central, 
tensor and spin-orbit components. We have shown that the magnitude and 
the shape of each of the internucleon correlation functions is clearly 
understood from the characteristics of the underlying two-nucleon interaction. 
The correlation function for the two-nucleon relative kinetic energy 
also indicates the importance of the tensor correlation 
and the short-range central repulsion in its momentum dependence. 

We have discussed the asymptotic behavior of the internucleon 
correlation functions. For a large separation of the two nucleons, the 
correlation functions are determined by negative energy solutions of 
three-body systems interacting via Coulomb potentials.  

Studying the internucleon correlation functions for heavier nuclei 
will be interesting and important because they give us direct
information on the distribution of the internucleon motion. 
We expect that the shapes of the functions do 
not differ drastically from those in the lightest nuclei, 
but the heights of the peaks are expected to be smaller, 
and the larger the mass number, the larger the spatial extension of the 
correlation function.

\appendix

\section{Correlation function for spin-orbit force}
\label{appendix.a}

In this appendix we show a method of calculating the $i$CF for 
the spin-orbit force following the formulation of Ref.~\cite{fbs}. 
We use the notation used there to be consistent with the formulation, so 
that a reader is referred to Ref.~\cite{fbs} for details. Since the 
matrix elements of the spin and isospin parts are calculated using 
a standard method, we here focus on the spatial part only. 
  
The spatial part of the spin-orbit force has the form $V(|{\vi
r}_{ij}|)({\vi r}_{ij}\times{\vi \pi}_{ij})$. We can express 
${\vi r}_{ij}={\vi r}_i-{\vi r}_j$
in terms of a linear combination of the relative 
coordinates ${\vi x}_k$, that is 
${\vi r}_{ij}=\sum_{k=1}^{N-1}w_k{\vi x}_k=\widetilde{w}{\vi x}$. 
Similarly ${\vi \pi}_{ij}$ is expressed as 
${\vi \pi}_{ij}=\sum_{k=1}^{N-1}\xi_k{\vi \pi}_k=\widetilde{\xi}{\vi
\pi}$, where ${\vi \pi}_k=-i\partial/\partial {\vi x}_k$ 
is a momentum operator conjugate to 
${\vi x}_k$. Thus  the spin-orbit 
force $V(|{\vi r}_{ij}|)({\vi r}_{ij}\times{\vi \pi}_{ij})$ is expressed 
as $V(\vert\widetilde{w}{\vi x}\vert) 
(\widetilde{w}{\vi x}\times \widetilde{\xi}{\vi \pi})$. 

The spin-orbit matrix element is calculated from the following expression 
\begin{eqnarray}
&& \langle F_{(L_3L_4)L'M'}({u_3},{u_4},A',{\vi x})\vert
V(\vert\widetilde{w}{\vi x}\vert) 
(\widetilde{w}{\vi x}\times \widetilde{\xi}{\vi \pi})_m
\vert F_{(L_1L_2)LM}(u_1,u_2,A,{\vi x})\rangle 
\nonumber \\
& &=\! \left(\prod_{i=1}^{4}{\frac{B_{L_i}}{L_i!}}
\int d{\vi e}_i \right) 
([Y_{L_3}({\vi e}_3) Y_{L_4}({\vi e}_4)]_{L'M'})^{*}
[Y_{L_1}({\vi e}_1)  Y_{L_2}({\vi e}_2)]_{LM} 
\left( \prod_{i=1}^4 
{\frac{\partial^{L_i}}{\partial{{\lambda}_i}^{L_i}}}\right)
\nonumber 
\\
&&\, \times\, 
\langle g({\lambda}_3{\vi e}_3 u_3\!+\!{\lambda}_4{\vi e}_4 u_4; A',{\vi x})
\vert V(\vert\widetilde{w}{\vi x}\vert) 
(\widetilde{w}{\vi x}\times \widetilde{\xi}{\vi \pi})_m
\vert g({\lambda}_1{\vi e}_1 u_1\!+\!{\lambda}_2{\vi e}_2 u_2; A,{\vi x})
\rangle\Big|_{\lambda_i=0},
\label{spme.formula}
\end{eqnarray}
where $({\vi a}\times {\vi b})_m$ stands for $-\sqrt{2}i[{\vi a}\times{\vi
b}]_{1m}\!=\!-(4\sqrt{2}\pi/3)iab[Y_1(\hat{\vi a})Y_1(\hat{\vi
b})]_{1m}$. 
To calculate the spin-orbit matrix element between the 
generating functions, we make use of the relation 
\begin{eqnarray}
& &\langle g({\lambda}_3{\vi e}_3 u_3\!+\!{\lambda}_4{\vi e}_4 u_4; A',{\vi x})
\vert V(\vert\widetilde{w}{\vi x}\vert) 
(\widetilde{w}{\vi x}\times \widetilde{\xi}{\vi \pi})_m
\vert g({\lambda}_1{\vi e}_1 u_1\!+\!{\lambda}_2{\vi e}_2 u_2; A,{\vi x})\rangle 
\nonumber \\
&=& \int d{\vi r}\, V(r) 
\langle g({\lambda}_3 u_3\!+\!{\lambda}_4{\vi e}_4 u_4; A',{\vi x})
\vert \delta(\widetilde{w}{\vi x}-{\vi r}) (\widetilde{w}{\vi x}\times
\widetilde{\xi}{\vi \pi})_m\vert g({\lambda}_1{\vi e}_1
u_1\!+\!{\lambda}_2{\vi e}_2 u_2; A,{\vi x})\rangle,
\nonumber \\
\end{eqnarray}
with 
\begin{eqnarray}
& &\langle g({\lambda}_3{\vi e}_3 u_3\!+\!{\lambda}_4{\vi e}_4 u_4; A',{\vi x})
\vert \delta(\widetilde{w}{\vi x}-{\vi r}) (\widetilde{w}{\vi x}\times \widetilde{\xi}{\vi \pi})
\vert g({\lambda}_1{\vi e}_1 u_1\!+\!{\lambda}_2{\vi e}_2 u_2; A,{\vi x})
\rangle \nonumber \\
& & \quad = -i\left({\frac{(2\pi)^{N-2}c}{{\rm det}B}}\right)^{\frac{3}{2}}
{\vi r} \times (\widetilde{\xi}{\vi z}\!+\!c
\widetilde{\xi}AB^{-1}w\widetilde{w}B^{-1}{\vi v})
\nonumber \\
& & \quad \, \times \, 
{\rm exp}\left({\frac{1}{2}}\widetilde{\vi v}B^{-1}{\vi v}
\!-\!{\frac{1}{2}}c({\vi r}\!-\!\widetilde{w}B^{-1}{\vi v})^2\right),
\label{spin-orbit.generic}
\end{eqnarray}
where $B=A+A'$, $c=(\widetilde{w}B^{-1}w)^{-1}$, and 
${\vi z}=\sum_{i=1}^4\lambda_i{\vi e}_iA_iB^{-1}$ with
$A_1=A_2=A',\, A_3=A_4=-A $.
When the radial form of the spin-orbit potential is scalar, i.e. 
$V$ is a function of $r$, we may omit 
$c\widetilde{\xi}AB^{-1}w\widetilde{w}B^{-1}{\vi v}$ thanks to the relation
\begin{equation}
\int d{\vi r}\, V(r)({\vi r}\times {\vi a})\,  
{\rm exp}\left(-{\frac{1}{2}}c({\vi r}-{\vi a})^2\right)=0,
\end{equation}
which leads to 
\begin{eqnarray}
& &\langle g({\lambda}_3{\vi e}_3 u_3\!+\!{\lambda}_4{\vi e}_4 u_4; A',{\vi x})
\vert V(\vert\widetilde{w}{\vi x}\vert) 
(\widetilde{w}{\vi x}\times \widetilde{\xi}{\vi \pi})_m
\vert g({\lambda}_1{\vi e}_1 u_1\!+\!{\lambda}_2{\vi e}_2 u_2; A,{\vi x})
\rangle \nonumber \\
& &\longrightarrow -i\left({\frac{(2\pi)^{N-2}c}{{\rm
		    det}B}}\right)^{\frac{3}{2}} \int d{\vi r}\, V(r)
({\vi r} \times \widetilde{\xi}{\vi z})_m
\nonumber \\
& &\qquad \times 
{\rm exp}\left(-{\frac{1}{2}}cr^2 +\sum_{j>i=1}^4\bar{\rho}_{ij}
\lambda_i\lambda_j {\vi e}_i\cdot{\vi e}_j 
+c\sum_{i=1}^4 \gamma_i\lambda_i {\vi e}_i\cdot{\vi r}\right),
\label{meso.gfn}
\end{eqnarray}
where $\gamma_i=\widetilde{w}B^{-1}u_i$,
$\bar{\rho}_{ij}=\widetilde{u_i}B^{-1}u_j-c\gamma_i\gamma_j$, and we set 
\begin{equation}
({\vi r} \times \widetilde{\xi}{\vi z})_m=-\sqrt{2}i \sum_{\alpha=1}^4T_{\alpha}
\lambda_{\alpha}[{\vi r}\times {\vi e}_{\alpha}]_{1m}
\label{def.tcoef}
\end{equation}
with
\begin{equation}
T_{\alpha}=\widetilde{\xi}A_{\alpha}B^{-1}u_{\alpha}.
\end{equation}
The symbol $\longrightarrow$ in Eq.~(\ref{meso.gfn}) indicates that the 
$\lambda_i^2$ terms in the exponent, which give no contribution 
to the required matrix element, are dropped. See Ref.~\cite{fbs} for details.  
Substitution of Eqs.~(\ref{meso.gfn}) and (\ref{def.tcoef}) into Eq.~(\ref{spme.formula}) 
yields a basic equation to obtain the spin-orbit matrix element
\begin{eqnarray}
&& \langle F_{(L_3L_4)L'M'}({u_3},{u_4},A',{\vi x})\vert
V(\vert\widetilde{w}{\vi x}\vert) 
(\widetilde{w}{\vi x}\times \widetilde{\xi}{\vi \pi})_m
\vert F_{(L_1L_2)LM}(u_1,u_2,A,{\vi x})\rangle 
\nonumber \\
& &=-\sqrt{2} \left({\frac{(2\pi)^{N-2}c}{{\rm det}B}}\right)^{\frac{3}{2}}
\sum_{\alpha=1}^4T_{\alpha}\int d{\vi r}\, V(r) \,{\rm e}^{-\frac{1}{2}cr^2}
\nonumber \\
& &\,\times \prod_{i=1}^{4}\left({\frac{B_{L_i}}{L_i!}}
\int d{\vi e}_i  \right) 
([Y_{L_3}({\vi e}_3) Y_{L_4}({\vi e}_4)]_{L'M'})^{*}
[Y_{L_1}({\vi e}_1)  Y_{L_2}({\vi e}_2)]_{LM} 
\nonumber \\
&&\, \times\, {\frac{\partial^{L_i}}{\partial{{\lambda}_i}^{L_i}}}
\lambda_{\alpha}[{\vi r}\times {\vi e}_{\alpha}]_{1m}
\,{\rm exp}\left(\sum_{j>i=1}^4\bar{\rho}_{ij}
\lambda_i\lambda_j {\vi e}_i\cdot{\vi e}_j 
+c\sum_{i=1}^4 \gamma_i\lambda_i {\vi e}_i\cdot{\vi r}\right)\Big|_{\lambda_i=0}.
\label{sp.gndformula}
\end{eqnarray}

We now have three expressions depending on ${\vi e}_i$, two in the 
exponent and one in $\lambda_{\alpha}[{\vi r}\times {\vi e}_{\alpha}]_{1m}$. 
Using the formula~(B.4) of Ref.~\cite{fbs}, we can rewrite the 
first expression as 
\begin{eqnarray}
& &{\rm exp}\left(\sum_{j>i=1}^4\bar{\rho}_{ij}
\lambda_i\lambda_j {\vi e}_i\cdot{\vi e}_j \right) 
\nonumber \\
& &\Rightarrow 
\sum_{p_{ij}}
\left(\prod_{j>i=1}^4\frac{(-1)^{p_{ij}}\sqrt{2p_{ij}+1}}{B_{p_{ij}}}(\bar{\rho}_{ij})^{p_{ij}}
\right) \sum_{\kappa}X(p_{13}p_{14}p_{23}p_{24};\kappa)
\nonumber \\
& & \quad \times \,
Y(p_{12}p_{12}p_{34}p_{34}\, p_{13}\!+\!p_{14}\, p_{23}\!+\!p_{24}\,
p_{13}\!+\!p_{23}\, p_{14}\!+\!p_{24}\, 0\kappa \kappa 0;\kappa \kappa)
\nonumber \\
& & \quad \times \, \Big[\big[Y_{p_{12}+p_{13}+p_{14}}({\vi e}_1)
Y_{p_{12}+p_{23}+p_{24}}({\vi e}_2)\big]_{\kappa}
\big[Y_{p_{13}+p_{23}+p_{34}}({\vi e}_3)Y_{p_{14}+p_{24}+p_{34}}({\vi e}_4)
\big]_{\kappa}\Big]_{00}
\nonumber \\
& & \quad \times \, \lambda_1^{p_{12}+p_{13}+p_{14}}\lambda_2^{p_{12}+p_{23}+p_{24}}
\lambda_3^{p_{13}+p_{23}+p_{34}}\lambda_4^{p_{14}+p_{24}+p_{34}}.
\label{e.e-term}
\end{eqnarray}
The symbol $\Rightarrow$ indicates that the angular momentum coupling
must be made to its maximum value for each ${\vi e}_i$. See 
Ref.~\cite{fbs} for details. 
The coefficients $X$ and $Y$ are defined in Eqs.~(B.7) and (B.9) of
Ref.~\cite{fbs}. With the use of Eq.~(B.22) of 
Ref.~\cite{fbs}, the second term reduces to 
\begin{eqnarray}
& & {\rm exp}\left(c\sum_{i=1}^4 \gamma_i\lambda_i {\vi e}_i\cdot{\vi r}\right)
\nonumber \\
& &\Rightarrow \sum_{q_i}
\left(\prod_{i=1}^4 \frac{(-1)^{q_i}(c\gamma_i)^{q_i}}{B_{q_i}}
r^{q_i}\right)
\sum_{\mu \mu' \ell}\sqrt{2\ell+1}\, C(q_1q_2;\mu)\, C(q_3q_4;\mu')\, 
C(\mu \mu';\ell)
\nonumber \\
& & \quad \times \, \Big[\big[[Y_{q_1}({\vi e}_1)Y_{q_2}({\vi e}_2)]_{\mu}
[Y_{q_3}({\vi e}_3)Y_{q_4}({\vi e}_4)]_{\mu'}\big]_{\ell}
Y_{\ell}(\hat{\vi r})\Big]_{00}
\lambda_1^{q_1}\lambda_2^{q_2}\lambda_3^{q_3}\lambda_4^{q_4}.
\label{e.r-term}
\end{eqnarray}
Here $C$ is a coefficient which couples two spherical harmonics 
with the same argument
\begin{equation}
C(l_1l_2;l_3)=\sqrt{\frac{(2l_1\!+\!1)(2l_2\!+\!1)}{4\pi (2l_3\!+\!1)}}
\langle l_10l_20\vert l_30 \rangle.
\end{equation}
Note that $\ell$ in the above summation~(\ref{e.r-term}) 
must be 1  to get a nonvanishing 
contribution in Eq.~(\ref{sp.gndformula}) 
because the term $\lambda_{\alpha}[{\vi r}\times {\vi e}_{\alpha}]_{1m}$ 
behaves like a vector in ${\vi r}$, which is simply given as 
\begin{equation}
\lambda_{\alpha}[{\vi r}\times {\vi e}_{\alpha}]_{1m} 
= -\frac{4\pi}{3}r [Y_1({\vi e}_{\alpha})Y_{1}(\hat{\vi r})]_{1m}\lambda_{\alpha}.
\label{e.cross.r-term}
\end{equation}

From Eqs.~(\ref{e.r-term}) and (\ref{e.cross.r-term}) we have 
\begin{eqnarray}
& &\Big[\big[[Y_{q_1}({\vi e}_1)Y_{q_2}({\vi e}_2)]_{\mu}
[Y_{q_3}({\vi e}_3)Y_{q_4}({\vi e}_4)]_{\mu'}\big]_{1}
Y_{1}(\hat{\vi r})\Big]_{00}
\lambda_1^{q_1}\lambda_2^{q_2}\lambda_3^{q_3}\lambda_4^{q_4}
 [Y_1({\vi e}_{\alpha})Y_{1}(\hat{\vi r})]_{1m}\lambda_{\alpha}
\nonumber \\
& & \Rightarrow \frac{1}{4\sqrt{3}\pi}
\Big[\big[[Y_{q_1}({\vi e}_1)Y_{q_2}({\vi e}_2)]_{\mu}
[Y_{q_3}({\vi e}_3)Y_{q_4}({\vi e}_4)]_{\mu'}\big]_{1}
Y_{1}({\vi e}_{\alpha})\Big]_{1m}
\lambda_1^{q_1}\lambda_2^{q_2}\lambda_3^{q_3}\lambda_4^{q_4}\lambda_{\alpha},
\label{coupl.2.3}
\end{eqnarray}
where the two $Y_1(\hat{\vi r})$'s are coupled to a scalar. Note that the 
integration over $\hat{\vi r}$ in Eq.~(\ref{sp.gndformula}) gives $4\pi$. 
It is convenient to express Eq.~(\ref{coupl.2.3}) as 
\begin{eqnarray} 
& &\Big[\big[[Y_{q_1}({\vi e}_1)Y_{q_2}({\vi e}_2)]_{\mu}
[Y_{q_3}({\vi e}_3)Y_{q_4}({\vi e}_4)]_{\mu'}\big]_{1}
Y_{1}({\vi e}_{\alpha})\Big]_{1m}
\lambda_1^{q_1}\lambda_2^{q_2}\lambda_3^{q_3}\lambda_4^{q_4}\lambda_{\alpha}
\nonumber \\
& &\Rightarrow \sum_{\nu \nu'}
K_{\alpha}(q_1 q_2 q_3 q_4 \mu \mu';\nu \nu')
\nonumber \\
& &\quad \times 
\big[[Y_{q_{1\alpha}}({\vi e}_1)Y_{q_{2\alpha}}({\vi e}_2)]_{\nu}
[Y_{q_{3\alpha}}({\vi e}_3)Y_{q_{4\alpha}}({\vi e}_4)]_{\nu'}\big]_{1m}
\lambda_1^{q_{1\alpha}}\lambda_2^{q_{2\alpha}}\lambda_3^{q_{3\alpha}}
\lambda_4^{q_{4\alpha}},
\label{redef.coupl.2.2}
\end{eqnarray}
where 
\begin{equation}
q_{i\alpha}=q_i+\delta_{i\alpha}.
\end{equation}
An expression for the coefficient $K_{\alpha}$ will be given later. 
The coupling of Eqs.~(\ref{e.e-term}) and (\ref{redef.coupl.2.2}) must lead to 
$\big[[Y_{L_1}({\vi e}_1)Y_{L_2}({\vi e}_2)]_{L}
[Y_{L_3}({\vi e}_3)Y_{L_4}({\vi e}_4)]_{L'}\big]_{1m}$ in order to have a nonvanishing 
contribution in Eq.~(\ref{sp.gndformula}). This coupling gives the following 
factor
\begin{eqnarray}
& &\Big[\big[Y_{p_{12}+p_{13}+p_{14}}({\vi e}_1)
Y_{p_{12}+p_{23}+p_{24}}({\vi e}_2)\big]_{\kappa}
\big[Y_{p_{13}+p_{23}+p_{34}}({\vi e}_3)Y_{p_{14}+p_{24}+p_{34}}({\vi e}_4)
\big]_{\kappa}\Big]_{00}
\nonumber \\
& &\, \times \big[[Y_{q_{1\alpha}}({\vi e}_1)Y_{q_{2\alpha}}({\vi e}_2)]_{\nu}
[Y_{q_{3\alpha}}({\vi e}_3)Y_{q_{4\alpha}}({\vi e}_4)]_{\nu'}\big]_{1m}
\nonumber \\
& &\Rightarrow Y(p_{12}\!+\!p_{13}\!+\!p_{14}\, p_{12}\!+\!p_{23}\!+\!p_{24}\, 
p_{13}\!+\!p_{23}\!+\!p_{34}\,  p_{14}\!+\!p_{24}\!+\!p_{34}\, 
q_{1\alpha} q_{2\alpha} q_{3\alpha} q_{4\alpha}\, \kappa \nu \nu' 
1; LL')
\nonumber \\
& & \quad \times \big[[Y_{L_1}({\vi e}_1)Y_{L_2}({\vi e}_2)]_{L}
[Y_{L_3}({\vi e}_3)Y_{L_4}({\vi e}_4)]_{L'}\big]_{1m}.
\end{eqnarray}
The values of $p_{ij}$ and $q_i$ must satisfy the following equations:
\begin{eqnarray}
& &p_{12}+p_{13}+p_{14}+q_{1\alpha}=L_1,\ \ \  
p_{12}+p_{23}+p_{24}+q_{2\alpha}=L_2, \nonumber \\
& &p_{13}+p_{23}+p_{34}+q_{3\alpha}=L_3,\ \ \  
p_{14}+p_{24}+p_{34}+q_{4\alpha}=L_4.
\end{eqnarray}
The operation prescribed in Eq.~(\ref{sp.gndformula}) is now easily performed. 

To sum up these results, we obtain the following formula: 
\begin{eqnarray}
&& \langle F_{(L_3L_4)L'M'}({u_3},{u_4},A',{\vi x})\vert
V(\vert\widetilde{w}{\vi x}\vert) 
(\widetilde{w}{\vi x}\times \widetilde{\xi}{\vi \pi})_m
\vert F_{(L_1L_2)LM}(u_1,u_2,A,{\vi x})\rangle 
\nonumber \\
& &= 4\pi \sqrt{\frac{2}{3}} \frac{(-1)^{L_1+L_2+L+L'+1}}{\sqrt{2L'+1}}
\langle LM 1m| L'M' \rangle \left(\prod_{i=1}^4B_{L_i}\right)
\left({\frac{(2\pi)^{N-2}c}{{\rm det}B}}\right)^{\frac{3}{2}}
\nonumber \\ 
& &\quad \times 
\sum_{\alpha=1}^4T_{\alpha}\sum_{p_{ij}}
\left(\prod_{j>i=1}^4\frac{(-1)^{p_{ij}}\sqrt{2p_{ij}+1}}{B_{p_{ij}}}(\bar{\rho}_{ij})^{p_{ij}}
\right) \sum_{\kappa}X(p_{13}p_{14}p_{23}p_{24};\kappa)
\nonumber \\
& & \quad \times \,
Y(p_{12}p_{12}p_{34}p_{34}\, p_{13}\!+\!p_{14}\, p_{23}\!+\!p_{24}\,
p_{13}\!+\!p_{23}\, p_{14}\!+\!p_{24}\, 0\kappa \kappa 0;\kappa \kappa)
\nonumber \\
& & \quad \times 
\sum_{q_i}
\left(\prod_{i=1}^4 \frac{(-1)^{q_i}(c\gamma_i)^{q_i}}{B_{q_i}}\right)
\sum_{\mu \mu' }\, C(q_1q_2;\mu)\, C(q_3q_4;\mu')\, C(\mu \mu';1)
\nonumber \\
& & \quad \times 
\int_0^{\infty} d{ r}\, r^{q_1+q_2+q_3+q_4+3} \,{\rm e}^{-\frac{1}{2}cr^2}\, V(r)
\sum_{\nu \nu'}K_{\alpha}(q_1 q_2 q_3 q_4 \mu \mu';\nu \nu') 
\nonumber \\
& & \quad \times Y(L_1\!-\!q_{1\alpha}\, L_2\!-\!q_{2\alpha}\, 
L_3\!-\!q_{3\alpha}\,  L_4\!-\!q_{4\alpha}\, 
q_{1\alpha} q_{2\alpha} q_{3\alpha} q_{4\alpha}\, \kappa \nu \nu' 
1; LL').
\label{LS.gndformula}
\end{eqnarray}
The coefficients $K_{\alpha}$ are given as follows:
\begin{eqnarray}
& &K_1(q_1q_2q_3q_4\mu \mu';\nu \nu')=-\delta_{\nu' \mu'}
C(q_1 1;q_{11})U(\mu' \mu 11;1\nu)U(q_2q_1\nu 1;\mu q_{11}),
\nonumber \\
& &K_2(q_1q_2q_3q_4\mu \mu';\nu \nu')=(-)^{\mu+\nu}\delta_{\nu' \mu'}
C(q_2 1;q_{22})U(\mu' \mu 11;1\nu)U(q_1q_2\nu 1;\mu q_{22}),
\nonumber \\
& &K_3(q_1q_2q_3q_4\mu \mu';\nu \nu')=(-)^{\mu'+\nu'+1}\delta_{\nu \mu}
C(q_3 1;q_{33})U(\mu \mu' 11;1\nu')U(q_4q_3\nu' 1;\mu' q_{33}),
\nonumber \\
& &K_4(q_1q_2q_3q_4\mu \mu';\nu \nu')=\delta_{\nu \mu}
C(q_4 1;q_{44})U(\mu \mu' 11;1\nu')U(q_3q_4\nu' 1;\mu' q_{44}).
\end{eqnarray}  
The values of $\nu$ and $\nu'$ are constrained by triangular 
relations which come from the unitary Racah coefficients $U$. 
Choosing $V(\vert\widetilde{w}{\vi x}\vert)=
\delta(\vert\widetilde{w}{\vi x}\vert-r)/r^2$ in Eq.~(\ref{LS.gndformula}) 
leads to the $i$CF for the spin-orbit force.

\section{Asymptotics of internucleon correlation functions}
\label{appendix.b}

Here we discuss the asymptotic form of the $i$CF. Let ${\vi x}_1$ 
(instead of ${\vi r}_{12}$)  
denote the relative distance vector of nucleons 1 and 2, and ${\vi x}_2$ 
denote the coordinate of their center of mass relative to the 
center of mass of the rest of the nucleus, which is called a nucleus R, 
containing $N-2$ nucleons.
When two nucleons are separated far in distance from R, 
all the nuclear forces 
can be neglected, and only the Coulomb interactions among them remain. 
The Hamiltonian $H$ of the whole system thus reduces to
\begin{equation}
t_1+t_2+\frac{e^2}{x_1}P_{1\pi}P_{2\pi}+\frac{Z_{R}e^2}{|\frac{1}{2}{\vi x}_1+{\vi x}_2|}
P_{1\pi}+\frac{Z_{R}e^2}{|-\frac{1}{2}{\vi x}_1+{\vi x}_2|}P_{2\pi}+H_{R},
\label{asym.H}
\end{equation}
where $t_1$ and $t_2$ are the kinetic energies, 
$t_1=-(\hbar^2/m)\partial^2/\partial {\vi x}_1^2$, and 
$t_2=-(\hbar^2/2\mu)\partial^2/\partial {\vi x}_2^2$ with $\mu=2(N-2)m/N$.  
The charge of the nucleus R is $Z_{R}e$, and its internal Hamiltonian 
is denoted by $H_{R}$.  
Corresponding to this decomposition, the wave function 
$\Psi$ for large $x_1$ takes the form
\begin{equation}
\Psi =\sum_{LS_{12}T_{12}IJ_RT_R} K_{LS_{12}T_{12}IJ_{R}T_{R}}
[[\Phi_L({\vi x}_1,{\vi x}_2)\chi_{S_{12}}]_I \eta_{T_{12}}
\Psi_{J_{R}T_{R}}]_{JM_JTM_T},
\label{asym.wf}
\end{equation}
where $\Psi_{J_{R}T_{R}}$ is the normalized wave function of 
R with spin $J_{R}$ and isospin $T_{R}$, though it may not be always an
eigenstate of $H_R$. Let $E_{J_RT_R}$ be the 
energy expectation value which $\Psi_{J_{R}T_{R}}$ gives, 
$\langle\Psi_{J_{R}T_{R}}|H_R|\Psi_{J_{R}T_{R}}\rangle $. 
The spin and isospin states of nucleons 1 and 2 are represented by 
$\chi_{S_{12}}$ and $\eta_{T_{12}}$. The summation labels of 
Eq.~(\ref{asym.wf}) 
run over all possible angular momenta which satisfy the 
angular momentum couplings and the parity conservation 
as well as the Fermi statistics of nucleons 1 and 2.

Using the asymptotic forms, (\ref{asym.H}) and (\ref{asym.wf}), in 
$H\Psi=E\Psi$, 
we find that the wave function $\Phi_{L}$ satisfies the three-body 
equation with only Coulomb potentials 
\begin{equation}
\left( 
t_1+t_2+\frac{e^2}{x_1}P_{1\pi}P_{2\pi}+\frac{Z_{R}e^2}{|\frac{1}{2}{\vi x}_1+{\vi x}_2|}
P_{1\pi}+\frac{Z_{R}e^2}{|-\frac{1}{2}{\vi x}_1+{\vi x}_2|}P_{2\pi}
\right)\Phi_{LM_L}=(E-E_{J_RT_R}) \Phi_{LM_L}.
\label{approx.asym}
\end{equation}
The energy $E-E_{J_RT_R}$ is however negative in contrast to a usual case~\cite{papp}. 
Since $\Psi$ is a bound state wave function, $\Phi_{LM_L}$ must also be bound. 
A solution for large $x_1$ can be obtained by solving the above equation. 
Here we attempt to obtain an approximate solution by taking the leading 
term of the nucleon-R Coulomb potential.  

Assuming that $x_1$ is much larger than $x_2$ makes it possible to simplify  
Eq.~(\ref{approx.asym}) to  
\begin{equation}
\left(t_1+t_2+\frac{Ze^2}{x_1}\right)\Phi_{LM_L}=(E-E_{J_RT_R}) \Phi_{LM_L},
\end{equation}
with 
\begin{equation}
Z=P_{1\pi}P_{2\pi}+2Z_{R}(P_{1\pi}+P_{2\pi}).
\label{def.charge}
\end{equation}
Since the coordinates ${\vi x}_1$ and ${\vi x}_2$ are now 
decoupled, we find a solution of the type of $\Phi_{LM_L}=
[\psi_{\ell}({\vi x}_1)\phi_{\lambda}({\vi x}_2)]_{LM_L}$. 
The function $\phi_{\lambda}({\vi x}_2)$ must be bound, that is, 
the matrix element of $t_2$, 
$\langle \phi_{\lambda m_{\lambda}}|t_2|\phi_{\lambda m_{\lambda}} \rangle$, 
must become negative, the value of which is set equal to 
$-{\hbar^2 q^2}/{2\mu}$. This is possible only for $\lambda=0$. 
Thus $\Phi_{LM_L}$ turns out to be of form 
$[\psi_{L}({\vi x}_1)\phi_{0}({\vi x}_2)]_{LM_L}$. 
Then the radial 
part of $\psi_{L}=f_L(x_1)Y_{L}(\hat{\vi x}_1)$ 
satisfies the equation
\begin{equation}
\left[\frac{d^2}{dx_1^2}+\frac{2}{x_1}\frac{d}{dx_1}-\frac{L(L+1)}{x_1^2}-
\frac{2\kappa \eta}{x_1}-\kappa^2\right]f_L(x_1)=0,
\label{radial.eq}
\end{equation}
where 
\begin{equation}
\eta=mZe^2/2\hbar^2\kappa,\ \ \ 
\kappa=\sqrt{\frac{m}{\hbar^2}\Big(-E+E_{J_RT_R}-\frac{\hbar^2
q^2}{2\mu}\Big) }.
\label{eta.kappa}
\end{equation}
This equation is the same as a scattering equation by a Coulomb
potential but with negative energies. 
A solution $f_L(x_1)$ which decreases for large $x_1$ is given by 
\begin{equation}
f_L(x_1)=\frac{1}{\kappa x_1}W_{-{\eta},L+\frac{1}{2}}(2\kappa
 x_1).
\label{radialwf}
\end{equation}
Here $W$ is the Whittaker
function~\cite{whittaker}, which is given using the confluent
hypergeometric function $F$ by
\begin{eqnarray}
W_{a,b}(z)&=&\frac{\Gamma(-2b)}{\Gamma(\frac{1}{2}-b-a)}z^{b+\frac{1}{2}}
{\rm e}^{-\frac{z}{2}}F(b-a+\textstyle{\frac{1}{2}},2b+1;z)\nonumber \\
&+&\frac{\Gamma(2b)}{\Gamma(\frac{1}{2}+b-a)}z^{-b+\frac{1}{2}}
{\rm e}^{-\frac{z}{2}}F(-b-a+\textstyle{\frac{1}{2}},-2b+1;z).
\end{eqnarray}
Here $\Gamma$ is the Gamma function. 
From the asymptotic form of the Whittaker function for large $z$, we have 
\begin{equation}
f_L(x_1) 
\sim x_1^{-1-{\eta}}{\rm e}^{-\kappa x_1}.
\label{asym.wf.coordinate}
\end{equation}
Substituting Eqs.~(\ref{asym.wf}) and (\ref{radialwf}) 
to Eq.~(\ref{dens.2}) gives the asymptotic form of 
$C^{(ST)}_{\cal O}(r)$. 

The value of $\kappa$ depends on whether nucleons 1 and 2 are both 
neutrons, or protons, or a neutron and a proton as well as 
on the value of $q$. 
The case which gives a minimum value of $\kappa$ determines the 
behavior of $i$CF at large distances.  
 
The behavior of $C(k)$ for small values of $k$ is given by the Fourier 
transform of the right side of  Eq.~(\ref{asym.wf.coordinate}), which reads 
\begin{eqnarray}
& &\frac{1}{(2\pi)^{3/2}} \int d{\vi x}_1\, {\rm e}^{-i{\vis k}\cdot{\vis x}_1} 
x_1^{-1-{\eta}} {\rm e}^{-\kappa x_1}Y_{L}({\hat{\vi x}_1})
\nonumber \\
& &\sim \frac{k^L}{\kappa^{L+\eta}
(k^2+\kappa^2)^{1-\eta}}
F\Big(\frac{L+1+\eta}{2},\frac{L+\eta}{2};L+\frac{3}{2};-\frac{k^2}{\kappa^2}
\Big)Y_L(\hat{\vi k}),
\end{eqnarray}
where $F$ is the Gauss hypergeometric series. 
The result is thus obtained as 
\begin{equation}
C(k) \sim \frac{k^{2L}}{(k^2+\kappa^2)^{2-2\eta}}
\left[F\Big(\frac{L+1+\eta}{2},\frac{L+\eta}{2};L+\frac{3}{2};-\frac{k^2}{\kappa^2}\Big)\right]^2.
\label{asym.c(k)}
\end{equation}

\bigskip
\vspace{1cm}
We are grateful to R. G. Lovas for his careful reading of the
manuscript and suggestions. 
We thank H. Feldmeier, T. Neff, M. Matsuo, T. Nakatsukasa and L. Tomio 
for useful discussions.


\begin{thebibliography}{99}
\bibitem{hk}P. Hohenberg and W. Kohn, Phys. Rev. {136} (1964) B864.
\bibitem{ks}W. Kohn and L.J. Sham, Phys. Rev. {140} (1965) A1133.
\bibitem{engel}J. Engel, Phys. Rev. C {75} 	(2007) 014306. 
\bibitem{barnea} N. Barnea, Phys. Rev. C {76}	(2007) 067302. 
\bibitem{gjb}B.G. Giraud, B.K. Jennings, and B.R. Barrett,
	Phys. Rev. A {78} (2008) 032507. 
\bibitem{forest}J.L. Forest, V.R. Pandharipande, S.C. Pieper,
	R.B. Wiringa, R. Schiavilla, and A. Arriaga, Phys. Rev. C {54} (1996) 646.
\bibitem{be8} R.B. Wiringa, S.C. Pieper, J. Carlson, and
	V.R. Pandharipande, Phys. Rev. C {62} (2000) 014001.
\bibitem{he4} W. Horiuchi and Y. Suzuki, Phys. Rev. C {78}	(2008) 034305.
\bibitem{ziesche}P. Ziesche, Int. J. Quantum Chem. {60}	(1996) 1361.
\bibitem{hetenyi} B. Hetenyi and S. Fantoni, Phys. Rev. Lett. {93}
	(2004) 	170202.
\bibitem{vmc} R.B. Wiringa, Phys. Rev. C {43} (1991) 1585; 
S.C. Pieper, R.B. Wiringa, and J. Carlson,  Phys. Rev. C {70} (2004)
054325.
\bibitem{bishop} R.F. Bishop, Theor. Chim. Acta {80} (1991) 95; 
R.F. Bishop, M.F. Flynn, M.C. Bosca, 
E. Buendia, and R. Guardiola, Phys. Rev. C {42} (1990)  
1341. 
\bibitem{chain}A. Fabrocini, F. Arias de Saavedra, G. Co', and
	P. Folgarait, Phys. Rev. C {57} (1998) 1668; 
A. Fabrocini, F. Arias de Saavedra, and G. Co', 
	Phys. Rev. C {61} (2000) 044302. 
\bibitem{alvioli} M. Alvioli, C. Ciofi degli Atti, and H. Morita,
	Phys. Rev. C {72} (2005) 054310.
\bibitem{ucom}H. Feldmeier, T. Neff, R. Roth, and J. Schnack,
	Nucl. Phys. {A 632} (1998) 61; T. Neff and H. Feldmeier, 
Nucl. Phys. {A 713} (2003) 311.
\bibitem{fbs} Y. Suzuki, W. Horiuchi, M. Orabi, and K. Arai, Few-Body Syst.  
{42} (2008) 33.
\bibitem{av8} R. B. Wiringa, V. G. J. Stoks, and R. Schiavilla, Phys. Rev. 
C\,{51} (1995) 38.
\bibitem{tamagaki} R. Tamagaki, Prog. Theor. Phys. {39} (1968) 91.
\bibitem{he6}W. Horiuchi and Y. Suzuki, Phys. Rev. C {76} (2007) 024311.
\bibitem{papp} Z. Papp, Phys. Rev. C {55} (1997) 1080.
\bibitem{whittaker}E.T. Whittaker and G.N. Watson, {\it A Course of
	Modern Analysis} (Cambridge Univ. Press, Cambridge, 1952).
\end{thebibliography}
\end{document}